\documentclass[prl,a4paper,twocolumn,showpacs,superscriptaddress,floatfix]{revtex4}
\usepackage{times}

\usepackage{graphicx}
\usepackage{amsmath}
\usepackage{amssymb}
\usepackage{bm}

\begin{document}

\title{Universal Braess Paradox in Open Quantum Dots}

\author{A.L.R. Barbosa}
\affiliation{
 Departamento de F\'isica, Universidade Federal Rural de Pernambuco, 52171-900 Recife - Pernambuco, Brazil}
\author{D. Bazeia}
\affiliation{Departamento de F\'isica, Universidade Federal da Para\'iba, 58051-970 Joa\~ao Pessoa  Para\'iba, Brazil}
\author{J.G.G.S. Ramos}
\affiliation{Departamento de F\'isica, Universidade Federal da Para\'iba, 58051-970 Joa\~ao Pessoa  Para\'iba, Brazil}

\date{\today}

\begin{abstract}

We present analytical and numerical results that demonstrate the presence of the Braess paradox in chaotic quantum dots. The paradox that we identify, originally perceived in classical networks, shows that the addition of more capacity to the network can suppress the current flow in the universal regime. We investigate the weak localization term, showing that it presents the paradox encoded in a saturation minimum of the conductance, provided the existence of hyper-flow in the external leads. In addition, we demonstrate that the weak localization suffers a transition signal depending on the over-capacity lead, and presents an echo on the magnetic crossover, before going to zero due to the full time reversal symmetry breaking. We also show that the quantum interference contribution can dominate the Ohm term in the presence of constrictions, and that the corresponding Fano factor engenders an anomalous behavior.

\end{abstract}


\maketitle

{\it Introduction.} - The Braess paradox asserts that the addition of a new road to the paths between two locations can counterintuitively increase the travel time of a vehicle \cite{DB}. The associated flux density depletion was also perceived in scenarios such as electrical networks \cite{pala}, wave packet propagation through a circular ring \cite{ufc}, mechanical devices, scanning gate microscopy among others \cite{Others}. Generically, we can say that the addition of extra capacity to a network can paradoxically lead to a depletion in its overall performance, under certain circunstances. 

The classic arguments for the Braess paradox include the Nash equilibrium condition \cite{Nash} about the competition between extra roads and the incentives to change the vehicle routes. In the electronic transport through a parallel network, Ohm's law imposes a conductance amplification according to the increasing of the number of parallel leads, i.e., the absence of Braess paradox in classical electronic circuits. On the other hand, the dynamics suggested by quantum mechanics impose, as in Nash dynamics, a peculiar complex competition between the subjacent wave phenomena on multi-terminal nanostructures \cite{Buttiker}. The Braess paradox has also been identified at the quantum level, both experimentally and numerically in an electrical network in \cite{pala}, and numerically in a circular quantum ring with the propagation of the wave packet calculated using the split-operator technique \cite{ufc}. The main conclusion is that the transport inefficiency also occurs at the nonometric scale, strongly influenced by quantum scattering and interference.

Such factual evidences motivate an investigation of quantum dots (QDs) coupled thought leads and performing appropriate networks. The more general investigation occurs in the universal regime, achieved when the chaos, due to the confinement of several resonances within each QD, generate statistical phenomena involving only fundamental symmetries of nature \cite{Caos1,Caos2}. The symmetries significantly affect both the interference, manifest in the quantum sector of the conductance (the weak localization term), and the corpuscular electronic nature, manifest in the noise power (the Fano factor) \cite{Noise}.

%%%%%%%%%%%%%%%%%%%%%%%%%%%%%%%%%%%%%%%%%%%
\begin{figure}[h!]
 \includegraphics[width=8cm,height=3.5cm]{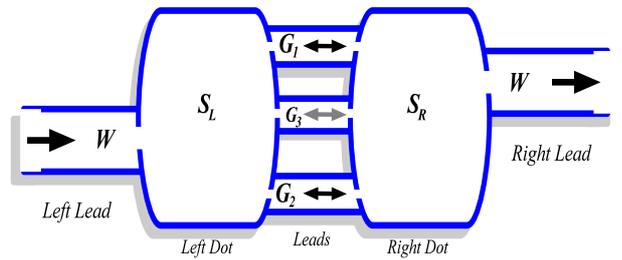}
\caption{\scriptsize(Color Online) The mesoscopic setup consists of two QDs, each one coupled to ideal electronic leads with an independent number of open propagating channels. The two QDs are coupled together with two leads, with the inclusion of a third, over-capacity lead.}\label{Fig1}
\end{figure}
%%%%%%%%%%%%%%%%%%%%%%%%%%%%%%%%%%%%%%%%%%%

\begin{figure}[t]
\centering
\includegraphics[width=8cm,height=6.5cm]{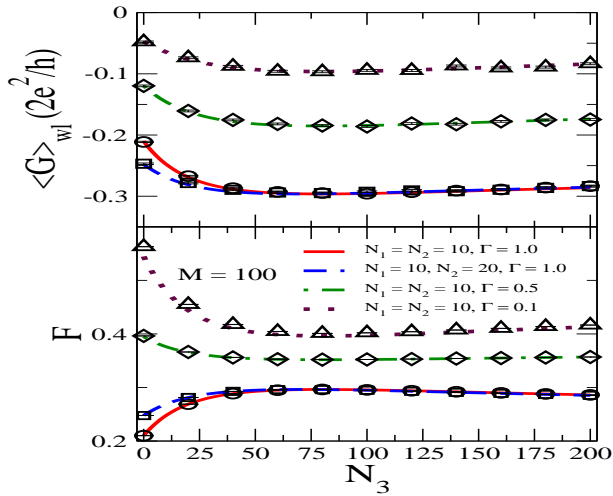}
\caption{\scriptsize (Color Online) The weak localization contribution (top panel) and the Fano factor (bottom panel, dimensionless), depicted in terms of the number of channels $N_3$ in the over-capacity lead $G_3$, for several values of $N_1$, $N_2$, and $\Gamma$. The curves represent the analytical results of Eqs.~(\ref{cond}) and (\ref{Fanomedia}), and the simbols correspond to the numerical simulation.}\label{Fig2}
\end{figure}

To investigate whether the Braess paradox appears in QDs, we deal with the mesoscopic setup depicted in Fig.~\eqref{Fig1}, which consists of two QDs, each one coupled to ideal electronic leads with an independent number of open propagating channels. The two QDs are coupled together by the two leads, and we add a third one, which is used to control the over-capacity of the apparatus. Intuitively, increasing the number of leads or channels between the QDs would amplify the quantum sector of the conductance. Nevertheless, the quantum nature of the phenomenon may imply a non Ohmic logic, giving rise to the Braess paradox at such mesoscopic level. In fact, we show that the conductance is consistent with the Braess paradox in mesoscopic nanostructures in the universal regime. The conclusion is founded on both analytical and numerical results, that match beautifully, as depicted in Fig.~\eqref{Fig2}.

{\it Scattering Approach} - We assume that the source and drain of electrons are coupled to the QDs by ideal leads, with $M_1$ (source) and $M_2$ (drain) open channels. The overall scattering matrix $S$ of the composite system is written as
\begin{equation}
S=\left(
\begin{array}{cc}
r & t \\
t' & r' \\
\end{array}\right),
\end{equation}
with $r \; (r')$ denoting a matrix of order $M_1$ ($M_2$) supporting the reflection amplitudes involving the open channels of the source (drain) coupled with the left (right) QD through ideal leads, while $t\; (t')$ is a $M_1 \times M_2$ ($M_2 \times M_1$) matrix buildt to contemplate the transmission amplitudes that connect the source and the drain channels.

The linear conductance of an open QD at zero temperature is given by the Landauer-B\"uttiker formula
\begin{equation}
G = \frac{2e^2}{h} g \;\; \textrm{with} \;\; g={\bf Tr}(tt^{\dagger}),
\end{equation}
where $g$ is the dimensionless conductance that depends of both the geometry of the QD and on external parameters \cite{PRL}. We can parameterize the internal scattering process between the two QDs, which connects the source channels to the drain channels, using the stub formalism \cite{brouwer,nos}. The scattering matrix is written in terms of a stationary portion $\bar{S}$ and a fluctuating portion $\delta S$, such that fluctuations in the scattering matrix are given by
\begin{eqnarray}
\delta \mathcal{S}=\mathcal{T}[1-\mathcal{U}\mathcal{R}(x)]^{-1}\mathcal{U}\mathcal{T}^\dagger.
\end{eqnarray}
The fluctuating portion of scattering matrix has dimension $2(M_1+N_T)\times 2(M_2+N_T)$, with $N_T=N_1+N_2+N_3$ denoting the total number of open channels of the $3$ leads that connect the two chaotic cavities. The block-diagonal matrix $\mathcal{U}$ represents the decomposition or projection, in the first block, of the source channels to the internal channels, and, in the second block, the internal channels to the drain channels \cite{Assis}. The transmission through the $3$ leads is described using the matrix $\mathcal{R}(x)$, while the contacts between outside leads and the modes of the chaotic cavities are described by $\mathcal{T}$ \cite{brouwer}. We introduce the following dimensionless parameters to characterize the intensity of time-reversal symmetry breaking in the system $x^2=h/\tau_B\Delta$, with $\tau_B$ and $\Delta$ denoting the magnetic decoherence time and the mean level spacing, respectively; see Ref.~\cite{nos}.

%%%%%%%%%%%%%%%%%%%%%%%%%%%%%%%%%%%
%%%%%%%%%%%%%%%%%%%%%%%%%%%%%

%%%%%%%%%%%%%%%%%%%%%%%%%%%%%
\begin{figure}[t]
 \includegraphics[width=8cm,height=4.5cm]{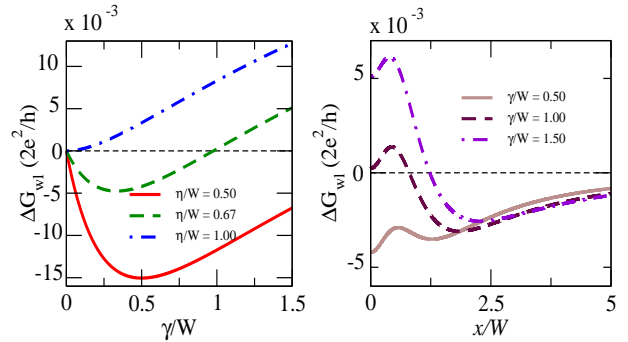}
\caption{\scriptsize (Color Online) In the left panel we depict $\Delta G_{wl}$ using the Eq.~(\ref{gwlideal}), in terms of the over-capacity open channels, $\gamma/W$. The analytical result indicates the presence of a minimum for $\eta/W<1$, associated to the constraint for critical congestion, $N_{3c}=W-N_1-N_2$. We depict the same quantity in the right panel, but now in terms of $x/W$. The curves clearly show the appearance of an echo; that is, the conductance increases and decreases abruptly, before going to zero with the increasing of the magnetic field. }\label{Fig3}
 \end{figure}
%%%%%%%%%%%%%%%%%%%%%%%%%%%%%

%%%%%%%%%%%%%%%%%%%%%%%%%%%%
\begin{figure*}[t]
\includegraphics[width=14cm,height=6cm]{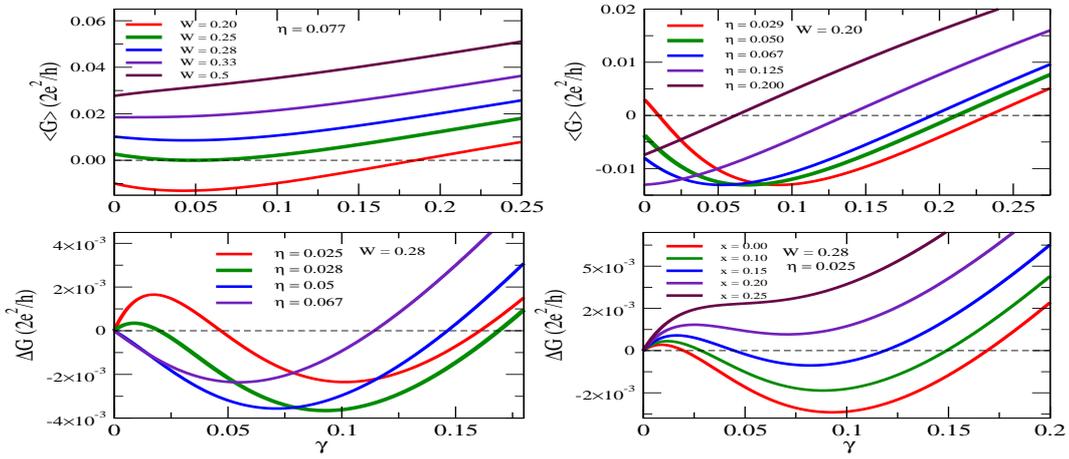}
\caption{\scriptsize (Color Online) Study of the competition between the Ohm law and the weak localization, in the opaque limit. In the top panels we depict $\left\langle G \right\rangle$, Eq.~(\ref{cond}), in terms of $\gamma=N_3 \Gamma_3$ for the over-capacity lead. In the left and right panels, we note that, for certain small values of $W$ and $\eta$, the weak localization overcomes the Ohm term. In the bottom panels we depict the difference $\Delta G= \left\langle G (\gamma)\right\rangle - \left\langle G (\gamma=0)\right\rangle$. In these panels we identify analytically an universal signal, shown in the minima of same magnitude, $10^{-3} 2e^2/h$. It engenders the Braess paradox of the experiment related in Ref.~\cite{pala} on a single measure. In particular, in the right panel we see that the magnetic crossover suppress the Braess paradox due to the breaking of the time-reversal symmetry.}  \label{Fig4}
\end{figure*}
%%%%%%%%%%%%%%%%%%%%%%%%%%%%%%

The ensemble average of Landauer-B\"{u}ttiker conductance transmission can be rewritten as a function of fluctuations of the scattering matrix
\begin{eqnarray}
\left\langle G \right\rangle =\frac{2e^2}{h} \left\langle g \right\rangle=\frac{2e^2}{h} \left\langle\textbf{ Tr} \left(\mathcal{P}_{L} \delta\mathcal{S} \mathcal{P}_{R} \delta\mathcal{S}^{\dagger}\right)\right\rangle, \label{landauer}
\end{eqnarray}
where $\mathcal{P}_{L}=diag(\textbf{1}_{W_{1}},\dots,0)$ and $\mathcal{P}_{R}=diag(0,\dots,\textbf{1}_{W_{2}})$ are projection matrices over the source and drain channels. For our purposes, we focus on a setup with an asymmetrical number of open channels in each one of the two leads, $N_1$ and $N_2$, respectively. The over-capacity lead of the setup, denoted by $3$, has $N_3$ open channels. We are now in a position to perform the standard diagrammatic method of integration over the unitary group \cite{brouwer,nos} to obtain the ensemble averages taking $M_1, M_2\gg 1$. Such diagrammatic analytical procedure renders the following simple expression for the transmission coefficient for the symmetric configuration with $W_1=W_2=W,$
\begin{widetext}
\begin{eqnarray}
\frac{\left\langle G\right\rangle}{2e^2/h}\! &=&\!\frac{W(\eta+\gamma)}{W+2\eta+2\gamma}
-\frac{W\left(\eta+\gamma\right)\left(W+\Gamma\left(\eta+
\gamma\right)\right)}{\left(W+2x^2\right)\left(W+2\eta+2\gamma\right)^2}-\frac{W\left(\Gamma(\eta+\gamma)^2-W(\eta+\gamma)+2W\left(G_1\Gamma_1\!+G_2\Gamma_2\!+\gamma\Gamma_3\right)\right)}{(W+2\eta+2\gamma+2x^2)(W+2\eta+2\gamma)^2},\;\;\label{cond}
\end{eqnarray}
\end{widetext}
where $W=M\Gamma$, $G_i=N_i\Gamma$  ($i=1,2,3$), $\eta= G_1+G_2$ and $\gamma=G_3$. Here, $\Gamma \in [0,1]$ describes the tunnel probability in the case of non-ideal contacts, which we also included in the analysis. The first term is the Ohm's Law, which has the usual behavior, and the next two terms gives the main quantum interference term, also known as weak localization, $\left\langle g\right\rangle_{wl}$. The crossover regime occurs for finite values of $x$, while the limit $x\longrightarrow \infty$ (large magnetic field strength) renders the pure unitary ensemble, for which $\left\langle g\right\rangle_{wl}$ vanishes due to the time-reversal symmetry breaking, as expected. In the symmetric system with a single lead between the chaotic cavities ($N_2=N_3=0$, $x=0$ and $W=G_1$ ) the  weak localization is $\left\langle g\right\rangle_{wl}=-2/9 \times (1/3+\Gamma)$, which reproduces the result of Ref.~\cite{GersonAssis}.

We can reach two relevant limits from our general result. Firstly, we consider the ideal configuration, with $\Gamma=\Gamma_1=\Gamma_2=\Gamma_3=1$. Here the weak localization term of Eq.~(\ref{cond}) simplifies to, after taking $x=0$,
\begin{eqnarray}
 \frac{\left\langle G\right\rangle_{wl}^{\textrm{ideal}}}{2e^2/h} &=&-2\frac{\left(W+\eta+\gamma\right)^2\left(\eta+\gamma\right)}{\left(W+2\eta+2\gamma\right)^3}.\label{gwlideal}
\end{eqnarray}
The second case is the opaque limit, defined as $N\longrightarrow \infty$ and $\Gamma\longrightarrow 0$, with $N\Gamma$ finite. Here the weak localization term of Eq.~(\ref{cond}) simplifies to, after also taking $x=0$,
\begin{eqnarray}
\frac{\left\langle G\right\rangle_{wl}^{\textrm{opaque}}}{2e^2/h} &=&-2\frac{W\left(\eta+\gamma\right)^2}{\left(W+2\eta+2\gamma\right)^3}.\label{gwlopaco}
\end{eqnarray}

{\it Braess Paradox} - The weak localization contribution presents the Braess paradox, as can be seen in Fig.~\eqref{Fig2}. Firstly, notice the presence of a minimum in both Eqs.~\eqref{gwlideal} and \eqref{gwlopaco} under the constraint $G_3=W-G_1-G_2>0$ for the over-capacity (OC)  extra lead. Such critical minimum $G_3^c$ is the same obtained in Ref.\cite{pala} for large quantum dots and indicates the condition of large transit outside, compared with the capacity of internal leads. We also consider constrictions, encoded on the tunneling barriers, and a large number of open channels so that congestion is generated. In fact, a backscattering process will be accessible on the nanostructure with the condition that its edges are in such a configuration that the number of channels of the inner conductor leads is less than the number of external channels, $G_3^c>0$. The graph depicted in Fig.~\eqref{Fig2} shows a clear depletion in the amplitude of the weak localization term and a corresponding minimum as a function of the open channels in the OC lead. The minimum is robust as in the experiment of Ref.~\cite{pala}, despite the total distinction between the universal Braess paradox we are reporting and the one of this experimental reference. Curiously, the minimum of the experiment occurs as the same constraint of the weak localization is satisfied.

In order to make the minimum evident in the ideal regime, in the left panel of Fig.~\eqref{Fig3} we plot $\Delta G_{wl}=\left\langle G (\gamma)\right\rangle_{wl} - \left\langle G (\gamma=0)\right\rangle_{wl}$ using the Eq.~\eqref{gwlideal} as a function of the OC channels, $\gamma/W$. The analytical result indicates the presence of a minimum for values of the number of channels in the OC lead for $\eta/W<1$, that is, for $N_1+N_2<W$, precisely associated with the critical hyper-traffic condition, $N_{3c}=W-N_1-N_2$. Moreover, depending on the configuration, the weak localization conductance term shows a signal transition, becoming positive after the minimum is reached. The magnetic crossover shows an anomalous behavior, also seen in Fig.~\eqref{Fig3}, where we plot $\Delta G_{wl} \times x$ in the right panel. The crossover magnetic field induces the presence of an echo, that is, before the weak location goes to zero with the magnetic field, it amplifies strongly with the variation of $x$.

We extend the investigation including the presence of finite barrier or constriction. Here we study the behavior of both Ohm and weak localization conductance terms. In the graphs of Fig.~\eqref{Fig4}, we depict the competition between the two terms in the opaque regime, plotting $\left\langle G \right\rangle$, given by Eq.~(\ref{cond}), as a function of $\gamma=N_3 \Gamma_3$ of the OC lead. The top panels of Fig.~\eqref{Fig4} show that for $W$ and $\eta$ sufficiently small, the weak localization term exceeds the Ohm term. This ensures that the Braess paradox exists in a very particular way in the universal regime, with strong competition between the two leading semiclassical terms. Induction of Braess paradox shows not only a minimum conductance, but also the signal inversion of the sum of the two terms. Also in the opaque regime, in the bottom panels of Fig.~\eqref{Fig4} one displays the difference $\Delta G=\left\langle G (\gamma)\right\rangle - \left\langle G (0)\right\rangle$, and there we observe analytically a universal signal in the minima of same magnitude, $10^{-3} 2e^2/h$, of the Braess paradox of the experiment of Ref.~\cite{pala}, which was realized for a single measure. In the right bottom panel, we see that the magnetic crossover suppresses the Braess paradox, due to the time-reversal symmetry breaking.

%%%%%%%%%%%%%%%%%%%%%%%%%%%%%%%%%%%%%%%%%%%%
\begin{figure}[t]
\includegraphics[width=7.6cm,height=7cm]{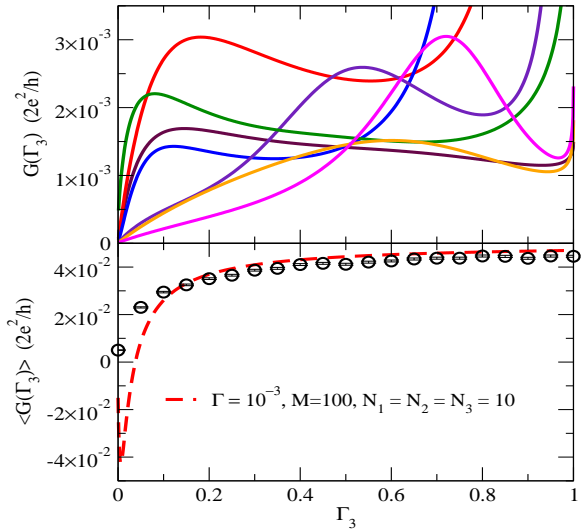}
\caption{\scriptsize (Color Online) The conductance is numerically depicted in the top panel, in the opaque regime, in terms of the tunneling probability,
$\Gamma_3$, of the over-capacity lead. Each curve represents a typical realization of two aleatory and independent QTs, and shows a clear non monotonic behavior. The curve depicted in the bottom panel refers to the analytical result of Eq.~(\ref{cond}), and the circles represent means over $10^5$ realizations, as the ones depicted in the top panel. The analytical result is compatible with the simulation
for $\Gamma_3>0.1$.} \label{Fig5}
\end{figure}
%%%%%%%%%%%%%%%%%%%%%%%%%%%%%%%%%%%%%%%%%%

The role of tunneling barrier becomes more evident in Fig.~\eqref{Fig5}, where we fix the remaining parameters and plot the conductance in the opaque regime as a function of the tunneling probability, $\Gamma_3$, of the OC lead. Each curve represents two individual realizations of random and independent quantum dots. Notice a clear non-monotonic behavior of the total conductance curves. However, despite the considerable universal fluctuations, the average is well behaved as can be seem in the lower curve of Fig.~\eqref{Fig5}, which represents the analytical result of Eq.~\eqref{cond}, with the simbols representing the mean over several realizations $(10^5)$, as the ones depicted in the top panel. The analytical results agree well with the simulation for $\Gamma_3>0.1$; for $\Gamma_3<0.1$ they differ, due to the fact that the high-order terms in the semi-classical expansion, taken into account in the analytical calculation, are supposed to have the same order of the Ohm contribution, thus competing with the weak localization in this regime. Such high-order terms appear due to the boundary effects in the quantum dots, and cannot be neglected in the numerical simulation.

The Fano factor was depicted in Fig.~\eqref{Fig2}. As one knows, it is defined as the reason between averages of shot noise and conductance 
\begin{equation}
F =\left\langle p\right\rangle/\left\langle g\right\rangle =1-\left\langle\textbf{Tr} \left(tt^{\dagger}\right)^2\right\rangle / \left\langle\textbf{Tr} \left(tt^{\dagger}\right)\right\rangle.
\end{equation}
Using again the diagrammatic method of Refs.~\cite{brouwer,nos} and taking the semiclassical limit, $M\gg 1$, we obtain for the Fano factor the following simple expression for ideal contacts
\begin{eqnarray}
 F &=& \frac{2\left(W+\eta+\gamma\right)^2\left(\eta+\gamma\right)}{\left(W+2\eta+2\gamma\right)^3}.\label{Fanomedia}
\end{eqnarray}
When we make $\eta=0$, and $\gamma=W$ in Eq. (\ref{Fanomedia}), we recover the results for two chaotic cavities in series, as it was obtained in Ref. \cite{GersonAssis}. Notice that the Fano factor has, curiously, the same absolute value of the weak localization term in the ideal case. However, the Fano factor presents an anomalous behavior, departing from the weak localization contribution in the presence of constrictions, as shown in Fig.~{\eqref{Fig2}}.

%%%%%%%%%%%%%%%%%%%%%%%%%%%%%%%
{\it Numerical Simulation} - We have performed a numerical simulation concerning the above investigation. We have followed the recent proposal
\cite{Assis}, to simulate networks of quantum dots. The basic ideia here is to modify the stub parametrization in an appropriate manner  and simulate each quantum dot through aleatory matrices, independent of the corresponding symmetry group. We have implemented the procedure always taking $10^5$ pairs of independent quantum dots in the ensemble to extract the mean values. The individuals realizations depicted in Fig.~\eqref{Fig5} also start from two independent quantum dots. The simulations confirm our analytical results, and they nicely describe the Fano factor depicted in Fig.~\eqref{Fig2}. 

%%%%%%%%%%%%%%%%%%%%%%%%%%%%%%%%%
{\it Summary and Conclusion} - In this Letter, we studied two quantum dots, coupled via three distinct leads. We showed that this simple network engenders an universal Braess paradox, which can be measured from distinct manners, with the weak localization conductance term in the ideal regime or in the presence of barriers, or in the competition between the Ohm and the weak localization terms. The investigation was extended to present analytical results in the crossover regime, where we could identify the appearance of an echo in the weak localization term.

The universal Braess paradox appears very clearly, in ensembles of pairs of QDs, after the critical hyper-flow condition $N_{3c}=W-N_1-N_2$ is reached. Moreover, we showed that it also shows up in specific realizations involving two QDs, thus presenting a behavior similar to the one related recently in the experiment of Ref.~\cite{pala}. The analytical and numerical results of the current study confirm each other, and bring the Braess paradox to such quantum environment, linking chaotic structures at the nanometric scale.

\begin{acknowledgments}
This work was partially supported by the Brazilian agencies CAPES, CNPq, and FACEPE.
\end{acknowledgments}

\end{document}